\documentstyle[aaspptwo]{article}

\newcommand{\hii}{H\,{\sc ii}\  }
\newcommand{\lam}{$\lambda$}
\newcommand{\oii}{[O\,{\sc ii}]}
\newcommand{\oiia}{\oii\,\lam3726}

\newcommand{\oiii}{[O\,{\sc iii}]}

\newcommand{\oiiib}{\oiii\,\lam5007}
\newcommand{\ha}{H$\alpha$}
\newcommand{\hb}{H$\beta$}
\newcommand{\ergs}{erg~s$^{-1}$}

\newcommand{\ergsa}{erg~s$^{-1}$~\AA$^{-1}$}
\newcommand{\ohb}{${\rm{I[O\,II}] + \rm{I[O\,III}]}\over{\rm{I(H}\beta)}$}
\newcommand{\msun}{$M_{\odot}$}
\newcommand{\zsun}{$Z_\odot$}
\newcommand{\philb}{$\phi\over L_B$\ }
\newcommand{\Mmin} {$M_{\rm{min}}$\ }
\newcommand{\Nmin} {$N_{\rm{min}}$\ }

\slugcomment{Accepted for publication in Astronomical Journal}

\begin{document}

\title { EMBEDDED CLUSTERS IN GIANT EXTRAGALACTIC \hii REGIONS: \\
 II. EVOLUTIONARY POPULATION SYNTHESIS MODEL }

\author {Y.\,D.~Mayya \altaffilmark{1} }
\affil{Indian Institute of Astrophysics, Bangalore 560 034 INDIA }
\affil{Infra Red Astronomy Group,
Tata Institute of Fundamental Research, Homi Bhabha Road, Colaba,
Bombay 400 005 INDIA}

\altaffiltext{1}
{Presently at the Bombay address ~~~E-mail: ydm@tifrvax.tifr.res.in}

\begin{abstract}

A stellar population synthesis model, suitable for comparison with Giant
Extragalactic \hii Regions (GEHRs), is constructed incorporating the recent
developments in modelling stellar evolution by Maeder and co-workers and
stellar
atmospheres by Kurucz. A number of quantities suitable for comparison with
broad band data of GEHRs in visible and near infrared parts of the spectrum are
synthesized in addition to the hydrogen and helium ionizing photon production
rates at solar metallicities, for three scenarios of star formation ---
(i) Instantaneous burst (IB) (ii) Continuous star formation (CSF) and (iii)
Two bursts of star formation, with the older burst rich in red supergiants.
For IB case, evolution of colors shows three distinct phases --- an initial
steady blue phase, followed by a red bump (5--15~Myr) and another steady
phase with colors intermediate to the earlier two phases. CSF colors
asymptotically reach peak values at $\sim 10$~Myr, never reaching the
reddest IB colors. Ionizing photon production rate falls off by an order
of magnitude in 6~Myr for IB, where as it almost remains constant for CSF
model. Two-burst models with burst separations $\sim 10$~Myr have properties of
both IB and CSF, simultaneously producing the red IB colors and high
ionizing photon rate, making such regions easily distinguishable using optical
observations. Flat IMFs result in bluest colors when the massive stars are
on the main sequence and reddest colors during the red supergiant phase of
the evolving massive stars. Errors on the computed quantities due to the
statistical uncertainties inherent in the process of star formation become
negligible for cluster masses in excess of $10^5$\,\msun.
Our GEHR spectra in the range 200~nm to 3~$\mu$m are found
to be in good agreement with the computations of Mas-Hesse and Kunth (1991).

\end{abstract}

\twocolumn

\section{Introduction}     

\noindent
External galaxies offer excellent opportunities for the study of star
formation under a variety of physical conditions. Many of the external galaxies
experience a much higher level of star formation activity compared to our own
Galaxy. Regions of current star formation are concentrated around the nucleus
in starburst and S\'ersic-Pastoriza galaxies (Balzano 1983, S\'ersic \&
Pastoriza 1965), while they are spread over the entire disks in blue compact
dwarfs, also known as \hii galaxies (Terlevich {\it et al.} 1991).
Giant extragalactic \hii regions scattered over the disks of
irregular galaxies and arms of spiral galaxies are the other regions
associated with intense on-going star formation activity. At the presently
achievable spatial resolutions, individual stars in most of the active
star-forming regions cannot be resolved or counted, owing to their greater
distances. Thus, information regarding the stellar population and the
star formation history in these regions has to be obtained
from the integrated properties of all the embedded stars.
Such a study would involve a comparison of synthesized properties of
star-forming regions with observables. Most of the early efforts in this
direction concentrated on synthesizing integrated colors of
galaxies (Larson \& Tinsley 1978,  Searle {\it et al.} 1973), which are
dominated by old low mass stars (age $\sim 10^{10}$ yr). Bruzual \& Charlot
(1993), Buzzoni (1989) synthesize integrated spectra of galaxies, which again
sample populations averaged over lifetimes of galaxies. The effect of active
star formation lasts at most for a few tens of millions of years, with
significant variations over period of fractions of million year (Myr). Thus
it is important to have time steps of the order of fractions of Myr to obtain
a complete picture of the early evolution of an active star-forming region.
Extensive grid of models covering a variety of observables
in selected bands from ultraviolet (UV) to near infrared (NIR) have been
synthesized by Rieke~{\it et al.} (1980), Lequeux~{\it et al.} (1981),
Melnick {\it et al.}~(1985), Olofsson~(1989),
Mas-Hesse \& Kunth (1991) and more recently by Leitherer \& Heckman (1995),
all suitable for young star forming complexes.

In all the above models a giant star-forming complex is assumed to consist
of a star cluster with an associated nebulosity around it. Ionizing photons
from hot OB stars of the star cluster are the source of ionization of the
nebula. Majority of the emission lines from the complex are produced in the
gaseous nebula, while the UV-optical continuum has contribution from both
stars and ionized gas. The dust particles embedded inside the \hii regions
absorb a fraction of ionizing photons, re-emitting them in far infrared
wavelengths. The models from different groups differ mainly in the input
data, and in the slightly different synthesis techniques used. There are two
main approaches adopted in the literature for synthesizing observable
quantities from star-forming regions. Bruzual \& Charlot (1993) and
Mas-Hesse \& Kunth (1991) among others have used observed stellar quantities
as input data base, while Buzzoni (1989) and Melnick {\it et al.}~(1985) have
used theoretical atmospheric models of Kurucz (1979) as input.
Although the former method is devoid of uncertainties in the model
atmospheres, it suffers from the incompleteness in the stellar input
data base, especially for the fast evolving evolutionary stages. In the
present work, we follow the second method, which gives the freedom to
study the active star-forming regions over a variety of physical conditions.
The main aim of the present model is to synthesize the spectra and
selected observable quantities in the optical and NIR regions,
incorporating the latest developments in modelling stellar evolution and
stellar atmospheres. Specifically the computations are done to compare the
observables in the optical bands obtained through CCD imaging. The most
important of these quantities is the ratio of \hb\ to
blue band luminosity. This ratio resembles the \hb\  equivalent width, but
unlike equivalent widths, it can be obtained accurately through imaging
observations. It should be noted that \hb\  equivalent widths are normally
obtained through slit spectroscopy, with slits often small compared to the
spatial extent of the giant star-forming complexes. This necessitates the
computations of quantities suitable for imaging observations.

We are considering the phase of the starburst in which the most recent one
or two bursts dominate the emission in all the bands from UV to NIR.
This phase corresponds to ages of the order of a few times $10^7$
yrs. We consider three scenarios of star formation, and evolve the
cluster up to 20~Myr after the epoch of star formation. In the first scenario,
all the cluster stars are assumed to have formed simultaneously (Instantaneous
Burst). Stars are assumed to be forming continuously (Continuous Star
Formation)
in the second scenario. Third scenario we considered has a second burst
occurring
in the star formation complex before the death of low to intermediate mass
stars from the earlier burst (multiple bursts).

In section 2, we define the model star-forming complex, and describe the
methods used for synthesizing the stellar quantities. The section also
includes the comparison with observed stellar quantities. The results of
the computations for the three scenarios discussed above is presented in
detail in section 3, while the effect of
statistical uncertainties inherent in the process of star formation are
discussed in section 4. The computed quantities are compared with the results
from earlier work in section 5. Main results are summarized in section 6.

In paper I (Mayya 1994) of this series we reported photometric data in
$BVR$\ha\ bands for around 180 \hii complexes in nine galaxies. In the third
and concluding paper in this series we will derive parameters related to star
formation by comparing the observational data of Paper~I with the synthesis
results of this paper.

\section{Description of Synthesis Technique}   

The present model assumes the star forming complex to be a star cluster
surrounded by an ionized medium. Its spectrum is thus
obtained by adding the spectrum of an ionized nebula to a pure cluster
spectrum. This section deals with the description of the synthesis technique
used in computing these two spectra.

\subsection{Definition of a Cluster}  

A star cluster in the model is defined by the following set of parameters : \\
1. Initial Mass Function (IMF) of stars defined by    \\
\hspace*{2cm} (a) Slope ($\alpha$) \\
\hspace*{2cm} (b) Upper cut-off ($m_u$) \\
\hspace*{2cm} (c) Lower cut-off ($m_l$) \\
2. Age ($t$) \\
3. Metallicity ($Z$) \\
4. Total Mass ($M_T$) \\
The interesting range of above parameters are,
$m_u\,=$\,30--120\msun, $\alpha\,=$\,1.0--3.5, $t\,=$\,0--20~Myr,
$Z\,=$\,0.001--0.04 ($Z\,=\,0.02$ corresponds to solar metallicity, \zsun).
Lower cut-off mass $m_l$ can have any value greater than 0.1\msun. $M_l$ is one
of the most difficult parameters to be evaluated, because of the dominance of
flux from higher mass stars in any wavelength region of observation. However,
a significant fraction of the total mass of star-forming regions is locked up
in low mass stars and hence the total mass as well as star formation rates are
critically dependent on $m_l$. Based on a number of observable quantities,
Rieke {\it et al.} (1980) found $m_l$ to be in the range 2--3\,\msun\ in
starburst regions. In the years that followed, there has been a number of
studies favoring $m_l$ values in the range 1--10\msun (see Scalo 1987 for a
discussion). We assume a value of 1\,\msun\ for $m_l$ throughout the present
work. Effect of changing $m_l$ on the computed quantities is discussed in
sec.~3.1. Following formulations are used in the computations of the spectral
energy distribution and luminosity in any band from a star cluster.
\begin{eqnarray}
L_\lambda (imf;t;Z) =& A \int\limits_{m_l}^{m_u}l_\lambda(m;t;Z)m^{-\alpha}
\,dm \\
\noalign{\hbox{and}}
L_{\rm B}(imf;t;Z) =& A \int\limits_{m_l}^{m_u}l_{\rm B}(m;t;Z)m^{-\alpha}
\,dm \\
\noalign{\hbox{with}}
          M_T =& A \int\limits_{m_l}^{m_u} m^{1-\alpha}dm \\
      l_{\lambda}(m;t;Z) =& \pi f_{\lambda}(m;t;Z)4\pi R^2 (m;t;Z)
\end{eqnarray}
where $f_{\lambda}(m;t;Z)$ is the intensity emitted from a star of mass
$m$ at an evolutionary stage $t$ at wavelength $\lambda$. $R(m;t;Z)$ is the
radius of the star, $A,$ the normalizing constant.
$Z$  is the abundance of the gas from which the stars are formed.
The stellar luminosity in any given band $B$ with a normalized response curve
$R_B(\lambda)$ is computed as,
\begin{equation}
 l_B(m;t;Z) = \int\limits_{band} l_{\lambda}(m;t;Z)R_B(\lambda)d\lambda.
\end{equation}
Lyman continuum photon rate is obtained by integrating the spectrum between
22.8 nm  and 91.2 nm. Thus,
\begin{equation}
 n_L(m;t;Z) = \int\limits_{22.8}^{91.2}l_{\lambda}(m;t;Z)
{ {\lambda}\over {\rm{hc}}}  d\lambda,
\end{equation}
with h and c being Planck's constant and speed of light respectively.
The above equations assume the cluster to be sufficiently massive so that
masses can be considered to be continuously populated, instead of
discretization into mass bins or spectral types. Under these conditions,
numerical integration can be performed rather than summing over all the
mass points. In sec. 4, we evaluate the minimum mass of the cluster above
which numerical integration rather than summation over mass is justifiable.

\subsection{Input Stellar Data}     

     The computations require a library of stellar spectra for a wide
range of masses at various evolutionary stages at the metallicities of
interest. We used the theoretically computed stellar spectra as input
dataset to our models. These spectra are synthesized by combining the
stellar evolutionary models from Maeder and co-workers and stellar atmospheric
models from Kurucz. We have used the evolutionary tracks from Schaller
{\it et al.} (1992), which is their most recent model for Z\,$=$\zsun. These
grids differ from the earlier grids from the same group (Maeder \& Meynet 1989,
Maeder 1990) mainly in having incorporated the new radiative opacities from
Rogers \& Iglesias~(1992). The models assume convective core
overshooting distance of 0.20 times the pressure scale height, and a
metallicity dependent mass loss rate (\.{\it M} $\sim Z^{0.5}$). For masses
greater than 20\msun, the authors have tabulated evolutionary results for
two values of the mass-loss rate, one being twice the other. The latter models
are computed with the intention of reproducing the properties of the low
luminosity Wolf-Rayet stars. As we are not aiming to count the number of
Wolf-Rayet stars as the cluster evolves, the two mass-loss schemes are
expected to give identical results. We have used the set
with lower mass-loss rates, which is more standard of the two.
Effective temperatures for Wolf-Rayet stars in these models have
been computed taking into account the extended atmospheres as described
in Kudritzki {\it et al.} (1989). Massive star evolution is carried out until
the end of core carbon burning in these models.

The atmospheric models of Kurucz~(1979) also have been recently updated
(Kurucz 1992). Apart from including many more layers than before to improve
the numerical accuracy of the computations, the new models have a resolution
of 1--2 nm in UV and optical regions of the spectrum. New computations are
available at 1221 wavelength points from 9 nm to 160 $\mu$m. The effective
temperature T$_{\rm{eff}}$ range of 3500--50000 has been covered for logarithm
surface gravity($\log g$) values of 0.0 to 5.0. Kurucz grids are adequate for
all stages during the evolution of stars in the mass range 1--120, except in
the following situations. Effective temperature of main sequence stars more
massive than 80\msun\ (earlier than O4) exceed 50000 K. These very hot models
do not differ much from blackbody and hence we used blackbody equations when
ever the temperature exceeded 50000 K. At the low temperature end, during the
fast evolving phase of red supergiants, the temperature drops below 3500 K,
for initial masses $m<20$ at ages $>10$~Myr. During this phase the
$\log g$ values also cross the lower limit of 0 in Kurucz grids. In these cases
we have used the 3500 K Kurucz models with $\log g\,=\,0$. This approximation
will not affect our computations for ages $<10$ Myr. Beyond this age, small
effects might be seen in the NIR.

\subsection{Computational details}   

Using Schaller {\it et al.}~(1992) evolutionary models, we have computed
$\log g$ and $T_{\rm{eff}}$ at the 51 tabulated evolutionary stages for
$Z\,=$\,\zsun\ and $1\leq m \leq 120$. The spectrum for a given $\log g$ and
$T_{\rm{eff}}$ is computed using Kurucz (1992) atmospheric models at $Z\,=$\,
\zsun, interpolating the spectrum linearly in $T_{\rm{eff}}$ and $\log g$ when
necessary. The absolute luminosities $l_\lambda$ are computed from these
spectra by using the mass-radius relationship as derived from the tables of
Schaller {\it et al.} (see eq. 4). This results in a spectrum sampled at all
the 1221 Kurucz wavelength points.

It can be seen from equations 1 and 2 that computations of cluster
quantities require a knowledge of $l_\lambda$ (or $l_{\rm B}$) at every mass
point chosen for integration. Schaller {\it et al.} tabulate the evolutionary
grids for 21 mass points in the mass range 0.8--120\msun, which is inadequate
for a complete integration over the chosen mass limits. Thus it is desirable
that the luminosities l$_\lambda$ are computed at sufficient number of mass
points to represent every phase during the evolution of the cluster.
This necessitated the generation of tracks for stellar masses intermediate to
the tabulated masses. After considerable experimentation we found it necessary
to extend the 21 mass Schaller {\it et al.} grids to 151 mass grids.
Of these 102 mass points lie between 10--40\msun, which is the most
important mass range for evolutions upto 20~Myr. Sampling interval is
smoothly increased from 0.2 to 0.5\msun\ in this mass range.
For smaller number of grids the evolutionary results are found to be
sensitive to the exact choice of the mass points.
The interpolation scheme adopted in generating the evolutionary tracks
for any intermediate mass is as follows. For a given mass, tracks for two
adjacent masses from Schaller {\it et al.} are identified. A track is
characterized by the mass, effective temperature ($T_{\rm{eff}}$) and
bolometric luminosity ($l_{\rm{bol}}$) and is sampled at 51 evolutionary
stages. New tracks for intermediate masses are generated by linear
interpolations in $\log(T_{\rm{eff}})$-mass and $\log(l_{\rm{bol}})$-mass
planes at all the 51 evolutionary stages. The tracks so obtained are sampled
at 0.1~Myr interval up to 20~Myr by linear interpolations in
$\log(T_{\rm{eff}})$-age and $\log(l_{\rm{bol}})$-age planes. The chosen
time step is suitable enough to represent all the short-term stellar
evolutionary phases. The band luminosities $l_B(m;t;Z)$ are further
interpolated at mass intervals of 0.05 dex for $m< 40$\msun\ and 0.2 dex for
higher masses to carry out the numerical integrations. Interpolations
are done linearly in $\log(l_B)$ - $\log(m)$ plane.

   The interpolation scheme described above is adopted after visually
verifying the linearity of the interpolated quantities in the plane of
interpolation.

\subsection{Synthesized Quantities and Comparison with Observations} 

Spectral energy distribution covering wavelengths in UV, optical and NIR
regions for any given stellar mass is one of the basic quantities
synthesized in our model. Luminosities in $BVRIJHKLM$ bands and the number
of hydrogen and helium ionizing photons are synthesized from these spectra
using the eqs in sec.~2.1. The response curves needed are taken from
Bessell~(1990) for $BVRI$
bands and Bessell \& Brett~(1988) for $JHKLM$ bands. Band luminosities are
transformed into colors using the zeropoints derived from the Vega spectrum of
Kurucz~(1992), assuming Vega to have colors 0.00 mag in all bands.
Zeropoints are also derived using the observed Vega spectrum for $BVRI$ bands
and model spectrum of Dreiling \& Bell~(1980) for $JHKLM$ bands. The
zeropoints derived from these two methods agree within 0.01 mag. Lyman
continuum photon rate is obtained by integrating model spectra shortward
of 91.2 nm as given in Eq.~6. Similarly the helium ionizing luminosity is
computed by integrating spectra shortward of 50.4 nm. The absolute luminosity
scale in the model is tied to Schaller {\it et al.} (1992) models.
This approach does not require the use of the highly uncertain
$T_{\rm{eff}}$-Bolometric Correction(BC) relation, to set the luminosity scale
of the model. On the other hand, we derive the BC over the region of interest
in T$_{\rm{eff}}$ and $\log g$ in order to compare with the observed values.

Before discussing the computations of cluster quantities, we compare the
computed model colors and BCs with observed values for different spectral
types and
luminosity classes. The most homogenous set of stellar photometric data in
$BVRIJHKL$ bands are due to Johnson~(1966). Flower~(1977) has updated the
T$_{\rm{eff}}$:$B-V$:BC scales making use of the most recent photometry
available at that time, and found Johnson (1966) scale to be in general
agreement in most cases, with $B-V$ colors redder by around 0.12 mag for
giants. For comparison with the model results we used $B-V$ and BC relations
with T$_{\rm{eff}}$ from Flower and the rest of the photometry from Johnson.
Johnson $V-R$ colors are transformed into Cousins system using the
transformation equations given by Bessell (1979) (after flipping the
mis-printed sign of the additive constant for the redder color range:
Bessell, private communication). There is more data available in the
literature for the evolving cool and hot phases of the stars (Humphreys \&
McElroy 1984; Elias {\it et al.} 1985), which are in general agreement
with the compilations of Johnson~(1966) and Flower (1977).

Lyman continuum photon production rate per unit surface area ($n_{\rm{lyman}}$)
from our computations is compared with that from Mas-Hesse \& Kunth (1991)
in Fig.~1(a). Other computed quantities such as bolometric correction (BC),
$B-V$, $V-R$ and $V-K$ colors are compared with the observed
values in Fig.~1(b)--(e) as a function of T$_{\rm{eff}}$. In each of these
figures, the observed values are denoted by filled circles for main sequence
stars and inverted triangles for supergiants. Main sequence model values are
shown by the solid lines, whereas the dashed line represents the supergiant
values as derived from our computations. It can be seen that on the whole,
there is good agreement between our model computations and observations. Our
T$_{\rm{eff}}$:$B-V$ scale for supergiants is around 0.03 mag bluer than that
of Flower (1977), with good agreement for main sequence stars. The differences
are of the order of 0.1 mag for red supergiants below 10000 K, where the
color-T$_{\rm{eff}}$ relation is very steep. 5--10\% errors in T$_{\rm{eff}}$
for a given color can easily explain these differences. $V-R$ and $V-K$ show
similar trends, with the model colors redder by as much as 0.2 and 0.5 mag
respectively for cool red supergiants. The observed BC for main
sequence and supergiants lie on either side of our computations, with the
differences of the order of 0.2 mag for T$_{\rm{eff}}$ greater than 10000 K.
For the coolest red supergiants the model BCs are higher (less negative)
by as much as 0.5 mag. As in the case of $B-V$, 5--10\% shift of model
T$_{\rm{eff}}$ scale towards the cooler side for a given color can explain the
observed differences in $V-R$, $V-K$ and BCs. Such shifts in T$_{\rm{eff}}$
are not unreasonable considering the different ways in which T$_{\rm{eff}}$
can be defined for extended atmospheres of stars, especially supergiants
(Baschek {\it et al.} 1991). The Lyman continuum rates agree within 0.05 dex
for hot main sequence stars, which mainly contribute to the total ionizing
photons in a cluster. The observed differences at lower temperatures where
the Lyman continuum steeply decreases with T$_{\rm{eff}}$ are thus not much
significant.

      In our model, the observational quantities are parameterized in terms of
mass and hence the absolute luminosities depend on the mass-luminosity
relation. As mentioned earlier, our scale is based on stellar evolutionary
models of Schaller {\it et al.} (1992), which are found to be in good agreement
with the values tabulated by Lang (1992) for main sequence stars.
Similar comparison for supergiants could not be performed because of the
uncertainty in the mass estimates for the model supergiants. However
model derived T$_{\rm{eff}}$-luminosity relation agrees with the
observed positions of red supergiants in the Hertzsprung-Russel diagram
(Humphreys \& McElroy 1984). Total Lyman continuum photon production rate for
high mass main sequence stars agrees
within 0.1 dex with the computations of Mas-Hesse \& Kunth (1991), Leitherer
(1990) and Panagia (1973). Our values begin to get systematically
different (lower) at masses lower than 40\msun, with differences reaching a
factor of 2 for 20\msun. The ionizing photon rate is lower by about a
factor of ten for these lower masses compared to those from the hottest
stars.

The basic stellar quantities and computed luminosities for the selected
stellar masses at zero age are given in Table~1(a) for solar metallicity.
Table~1(b) lists the same quantities during the red and blue
supergiant phases of the stars. The latter table contains quantities which
are obtained by taking the weighted averages over the supergiant phases.
The quantities tabulated in the two tables are:

\noindent
Mass, M$_{\rm{ms}}$ : Zero age main sequence (ZAMS) mass of the star \\
M$_{\rm{sg}}$ : Mass of the star during its supergiant phase \\
$T$ : Time spent in main sequence phase (Tab.~1a) and supergiant phases
(Tab.~1b) \\
T$_{\rm{eff}}$, BC, rad, M$_{\rm{bol}}$ : Effective temperature, Bolometric
Correction, radius and bolometric magnitude of stars \\
${M\over{L_V}}$ : Ratio of stellar mass to the luminosity in the visual
band, both expressed in solar units. M$_{\rm{ms}}$ is used as mass of the
supergiants rather than the highly mass-loss dependent M$_{\rm{sg}}$.\\
$N_L$ : Number of ionizing photons emitted from the star per second \\
$f_{\rm{He}}$ : Fraction of lyman continuum photons ($\lambda\le
91.2$~nm) available for the ionization of helium also ($\lambda\le 50.4$~nm) \\
\philb : Ratio of the expected \hb\ luminosity from the nebula surrounding
the star to the blue band luminosity. See sec.~3.1 for an exact definition
of this quantity. \philb, $f_{\rm{He}}$ and $N_L$ are meaningful only for
very hot stars (T$_{\rm{eff}} > 25000$K) \\
colors : Optical and NIR broad band colors. \\

\subsection{Computations of Nebular Flux}  

Ultraviolet photons from hot stars in the cluster ionize the
surrounding gas leaving hydrogen, helium, oxygen, nitrogen, carbon and
other astrophysically abundant elements in different ionized states.
For ionization bounded nebulae, all of these ionizing photons are used
up by hydrogen, and at ionization equilibrium, every ionization is followed
by recombinations to one of the levels in the hydrogen atom (Osterbrock 1989).
In thermal equilibrium, the energy input to the nebula in the form
of kinetic energy of the electrons is lost due to collisions and
radiative recombinations, leaving the nebula at a fixed temperature.
This heated gas along with the ions in
various excited states gives rise to a spectrum characteristic of \hii
regions. The basic physics governing the emission mechanism from these
regions is well understood and hence its theoretical spectrum
can be synthesized from only a few input parameters such as temperature,
density, abundance of elements and absolute flux in one of the Balmer lines.
In the next two subsections, we describe the method we have followed in
synthesizing the continuum and line spectrum of the nebula.

\subsubsection {Estimation of the Gaseous Continuum}  

The main emission mechanisms contributing to continuum emission from a gas
are free-free, free-bound, and two-photon ($2q$) emissions from hydrogen and
helium. The net emission coefficient $\gamma_{\rm{eff}}$ can be expressed as
a sum of these emission coefficients, weighted by the relative abundance of the
particular ionized state.
\begin{equation}
{{\gamma_{\rm{eff}}}} = \gamma {\rm {(H\,I)}} +
\gamma {\rm {(2q)}}  + {{n{\rm{(He}}^+ {\rm )}} \over {n{\rm{(H)}}}}
\gamma {\rm {(HeI)}} + {n{\rm{(He}}^{++}{\rm )} \over {n{\rm{(H)}}}}
\gamma {\rm {(HeII)}},
\end{equation}
where $n$(H) is the total number density of hydrogen atoms in both ionized
and neutral states. In our computations we assume a helium abundance of
10\% by number all of which is in the singly ionized state. Individual
$\gamma$'s as computed and tabulated by Brown \& Mathews (1970) are used here
for $\lambda <1\mu$m. For longer wavelengths computation of Ferland (1980) have
been used. The continuum of the gas spectrum in \ergsa\ is related to
$\gamma_{\rm{eff}}$ by,
\begin{equation}
 {{\rm L}^{\rm{neb}}_\lambda} = \gamma_{\rm{eff}}
\left( c\over {\lambda^2}\right)\left( 1\over {{\rm E}_{{\rm H}\beta}
\alpha^{\rm{eff}}_{{\rm H}\beta} ({\rm H}^0,T_e ) }\right)
{\rm L}_{{\rm H}\beta}.
\end{equation}
E$_{{\rm H}\beta}$ is the energy of the \hb\  photon,
$\alpha^{\rm{eff}}_{{\rm H}\beta} ({\rm H}^0,T_e )$ is the \hb\  recombination
coefficient and L$_{{\rm H}\beta}$ is the \hb\  line luminosity in \ergs.
$T_e$ is the temperature of the nebula at thermal equilibrium.

\subsubsection{Estimation of the Gaseous Emission line Strengths} 

Osterbrock (1974, 1989) has compiled the emission line strengths of various
transitions of hydrogen and helium relative to \hb\  line strength for typical
densities and temperatures of \hii regions. The strength of emission lines
from other elements however are dependent on the relative abundance of
the elements, the state of ionization, the temperature structure of the nebula
{\it etc.} and hence their computations require detailed modelling.
Such models have been constructed by Dopita \&
Evans (1986), Rubin (1985), McCall {\it et al.}~(1985) among others.
{}From observations of a large sample of extragalactic \hii regions McCall
{\it et al.}  have shown that the majority of the emission line
ratios can be parameterized by a single parameter, namely \ohb\ . The basic
physics behind this can be summarized as follows. The cooling of the
nebula is mainly controlled by the radiative losses in oxygen ions, which
constitute more than $45\%$ (Lequeux {\it et al.} 1979) of metallic mass.
This gives rise to an anti-correlation between oxygen abundance and nebular
temperature. Most of the oxygen in GEHRs is in singly (\oii) and doubly
(\oiii) ionized states, both of which have bright emission lines in the
optical region (\oiia\ and \oiiib). Hence, oxygen abundance can be quite
accurately
estimated using one or both of these lines. The observed correlations of line
ratios of other elements with oxygen abundance suggests that the relative
abundance of other elements such as C, N and S do not vary much from region
to region. Photoionization models are successful in reproducing the observed
correlation between line ratios and oxygen abundance, enabling the
prediction of line ratios from a knowledge of oxygen abundance alone.

Our interest here is to estimate the contribution of emission line
fluxes to $BVR$ band fluxes. We considered all the bright lines of N, O and S
apart from hydrogen lines as computed by McCall {\it et al.}~(1985). Their
tables 15 and 12 help us to estimate the oxygen abundance from the strengths
of \oiia\  and \oiiib\ lines, which in turn are used to estimate the strength
of various lines from other elements.

The nebular spectrum is then obtained by adding selected emission lines to the
continuum spectrum. Nebular spectrum when added to the cluster spectrum
results in a spectrum expected from a star-forming complex.

\section{Star Formation Scenarios} 

We consider three possible scenarios of star formation which might be at
work in GEHRs. In the simplest case referred as {\it Instantaneous Burst}
(IB), all the stars are assumed to have formed over a short time scale
(coeval) compared to the lifetime of even the most massive stars. The other
extreme case is to assume the star formation to be {\it continuous} over a
period of time with a star formation rate (SFR), which is either a
{\it constant} (CSFR) or {\it exponentially decreasing} (ESFR) function of
time. Thirdly we consider an intermediate scenario between the two mentioned.
We model the effect of two instantaneous bursts separated over time scales
comparable to the lifetime of the stars produced in the burst. Thus at any
given time, there will be stars from two generations. We refer to this model as
{\it Multiple Bursts} (MB). The results from
these three models are described separately in the following sections.

\subsection{Instantaneous Burst of Star Formation} 

The IB of star formation is fully described by the equations in sec.~2.1.
All the quantities synthesized for individual stars are also synthesized for
clusters. The dependence of the computed quantities on the IMF
parameters for a zero-age cluster is presented in Tab.~2(a). Model
numbers are given in the first column. The IMF
parameters, namely the upper cut-off mass and slope are given in columns 2
and 3. Ratios of the total mass to bolometric (${M\over L}$) and visual
(${M\over L_v}$) luminosities are given in columns 4 and 5 in solar units.
Lyman continuum photon production rate ($N_L$) per unit mass of the cluster
is given in column 6. Fraction of the hydrogen ionizing
photons which are also capable of ionizing helium are given in column 7.
Lyman continuum photon rate $N_L$ is also expressed in units of expected
\hb\  luminosity $\phi$ from a nebula at 10000 K and defined as,\\
\begin{equation}
\phi = 4.78\times 10^{-13} N_L.
\end{equation}
$\phi$ in the above equation has dimensions of erg\,s$^{-1}$\,\msun$^{-1}$,
and $N_L$ in ph\,s$^{-1}$\,\msun$^{-1}$. Column 8 contains \philb, which
is the ratio of the expected \hb\ to blue band luminosity. \hb\  equivalent
width is directly related to \philb\  as can be seen from Fig.~2. The
relationship is derived by evolving a cluster up to 20~Myr, with
the dots on the curves separated by 0.5 Myr interval.
Column 8 contains this ratio. Columns 9--14 contain optical and NIR
colors of the cluster (nebular contribution not taken into account). \hb\ and
\ha\ equivalent widths in angstrom are given in column 15 and 16 respectively,
assuming a gas temperature of 10000 K.

The effect of changing lower cut-off masses on the computed quantities is
illustrated in Tab.~2(b), expressed as differences from an IMF with
$m_l\,=\,1$\msun\  in logarithm or magnitude units. The tabulated values
correspond to an upper cut-off mass of 60\msun, but have identical values for
$30<m_u<120$\msun. Tabulated quantities in columns 4--16 are identical to
those in Tab.~1, except in column 7, which represents the factor by which
the mass of the cluster decreases due to the change in $m_l$ from a
value of 1\,\msun. It can be
seen that the quantities which are most affected by changing the lower cut-off
mass are $\log ({M\over L})$, $\log ({M\over {L_v}})$ and $N_L$. Changing
the lower cut-off has only the effect of changing the total mass of the
cluster with negligible change in the luminosities, as can be inferred
by comparing columns 4--6 with column 7. An IMF with a lower
cut-off of 1\msun\ contains 4--6 times more mass compared to an IMF with
$m_l\,=\,10$\msun\ for IMF slope$\,=$\,2.5 and 32--38 times more for
slope$\,=$\,3.5 for $m_u$ in the range 120 to 30\msun.
Among colors $V-K$ is most sensitive to changes in $m_l$ due to
non-negligible contribution to $K$ band luminosity from low mass stars. However
contribution from stars less massive than 1\,\msun\ is negligible to any
band luminosity and hence colors will remain the same. Quantities in
columns~4--6 can be reproduced for a different $m_l$ by adding the entries
in Tab.~2(a) to the logarithm of the quantity,
\begin{eqnarray}
f_M &= {{1 - m_u^{2-\alpha}} \over {m_l^{2-\alpha} - m_u^{2-\alpha}} }.
\end{eqnarray}
A real giant star-forming complex differs from its definition in the model in
the following ways. \\
1. Dust can be present inside star-forming complexes, which absorb and hence
reduce the total number of photons available for ionization. Throughout
their computations Mas-Hesse \& Kunth (1991) have assumed that 30\% of the
ionizing photons are absorbed by dust, based on the work of Belfort~{\it et
al.}
(1987).
However Mezger (1978) has shown this fraction to be highly uncertain having
values anywhere between 0.2 to 0.7 depending on the local conditions. Because
of the large uncertainty in the value, we prefer to present all the model
values without correcting for dust absorption. We leave it as a free parameter
to be assumed or determined while comparing with observations. \\
2. Reddening due to foreground interstellar dust and nebular continuum emission
in different bands make the observed colors different from the values computed
for a pure cluster. Both these effects can be handled in a better way using
available information on the physical conditions of the individual regions
while comparing with observations. Thus we tabulate the pure cluster values
instead of taking into the effects of interstellar extinction and nebular
emission in an average way. \\
3. Coevally formed stars take different amounts of time to reach the ZAMS. For
example a star of 100\msun\ takes only $1.9\times10^4$~yr, where as a star of
1\msun\ takes 75~Myr (Ezer \& Cameron 1967, Beech \& Mitalas 1994).
Schaller {\it et al.} (1992) models we have used considers the stellar
evolution from ZAMS stage onwards, thus our early stage cluster evolution
might be affected. Using the  pre-main sequence (PMS) evolutionary tracks of
Ezer \& Cameron (1967), we estimated the PMS contribution relative to the ZAMS
luminosities in various bands. PMS luminosity of even the low mass stars
remains within a factor of 2--3 of the ZAMS value for more than 80\% of
the PMS evolutionary stage. On the other hand, in the earlier stages,
cluster luminosity is dominated by the luminous massive main-sequence
stars, whose bolometric luminosity is $10^5--10^6$ times that of
1\,\msun\ star, rendering the inclusion of PMS stages unnecessary. Thus
computations of cluster quantities based on Schaller {\it et al.} models are
valid even at stages where low mass stars are yet to reach the ZAMS.

More massive a star is, earlier it exhausts the nuclear fuel and hence the
evolution of the IB model of SF is characterized by the selective removal of
massive luminous stars from the cluster. As the massive stars wander around
in the H-R diagram during their post main-sequence evolution, they affect the
integrated properties of the cluster, even when their number is small. We have
evolved the cluster at steps of 0.1~Myr interval upto 20~Myr, when the highest
surviving mass is $\sim 11$\msun. A detailed discussion of the these results
will be presented in sec.~3.4 along with the other two scenarios of star
formation considered.

\subsection{Continuous Star Formation}   

In the previous section we considered the evolution of a cluster
containing coeval stars formed in an instantaneous burst. Now consider
the case where stars are formed continuously resulting in stars of
different ages in the same cluster. This scenario, known as continuous
star formation (CSF), is characterized by a star formation rate (SFR),
the rate at which new stars are added to the cluster. We have considered
two kinds of SFRs, a constant rate of star formation (CSFR)
and an exponentially decreasing star formation rate (ESFR).

Computationally CSF scenario is realized by summing luminosities over all
surviving masses, which is mathematically represented as,
\begin{eqnarray}
L_{\rm
B}^{\rm{csf}}(imf;t;Z)=&{{A\,\int\limits_{0}^{t}\,\int\limits_{m_l}^{m_u}
\,l_{\rm B}(m;t^\prime;Z)\, m^{-\alpha}\, R(t-t^\prime) \,dm\,dt^\prime}
\over{ \int\limits_{0}^{t} R(t^\prime)\,dt^\prime}} \\
\noalign{\hbox{with}}
R(t^\prime) =& R_0  ~~~~~~~~~~~~~ {\rm{for~CSFR}} \\
\noalign{\hbox{and}}
R(t^\prime) =& R_0\,\exp({-{{t^\prime}\over{t_0}}})  ~~~~~~ {\rm{for~ESFR}}
\end{eqnarray}
$R(t^\prime)$ represents the rate of star formation at an epoch
$t^\prime$ with $R_0$ and $t_0$, the SFR at $t=0$ and the e-folding time
of SFR respectively. The denominator in eq.~11 is a measure of total mass in
all the stars and $A$ is the normalization constant defined in eq.~3.

\subsection{Multiple Bursts of Star Formation}  

We now consider the third scenario of star formation, in which a second burst
of star formation takes place before the death of intermediate--low mass
stars from the earlier burst. In several galactic OB associations, neighbouring
regions of star formation are found to differ in age by $\sim$10~Myr,
which is thought to be evidence for sequential star formation (Elmegreen
1992). The energetic events such as stellar winds, supernova activity,
typical of massive stars, might be responsible for the sequential
star formation (Elmegreen \& Lada 1977). In distant star forming regions, the
resolution may not be good enough to resolve the populations spatially,
and hence one has to infer the presence of more than one burst based on the
integrated light. In this study we aim at providing observational plots where
regions with two bursts of star formation can be identified.

Computations are done by adding the fluxes from two populations
of equal strength, one younger than 6~Myr and the other older than 6~Myr.
Each populations is assumed to be independently evolving like an IB
with identical IMFs.

\subsection{Discussion of Evolutionary Results} 

    As discussed in section 2.1, an IB cluster is defined by its age,
metallicity and total mass apart from the three IMF parameters.
CSF and MB scenarios have additional parameters defining them. Thus the total
parameter space involved is large for full presentation of the results. It
should also
be noted that Mas-Hesse \& Kunth (1991) and more recently Leitherer \&
Heckman (1995) have presented results for considerable section of the
parameter space. In this study, we concentrate on comparison of the three
scenarios of star formation for typical values of the parameters,
instead of presenting results over the entire parameter space.
We discuss the IB scenario in greater detail for two reasons --- (i) The
evolutionary trends in this case can be directly associated with the evolution
of stars and (ii) The results of IB scenario can be used to explain the
composite models such as CSF and MB.

   As has been pointed out earlier, we have fixed the metallicities at
$Z=$\zsun\ and normalized the quantities to cluster mass $M_T=1$\msun.
An IMF with $m_l\,=\,1, m_u\,=\,60,$ and slope$\,=\,2.5$, which has
slope close to Salpeter's value of 2.35 for the solar neighbourhood, is
chosen as a representative IMF in our computations. We refer this IMF as
``standard'' in rest of the article. The effects of
different IMF parameters are inferred by computing quantities for
two extreme IMFs --- (i) $m_l\,=\,1, m_u\,=\,30,$ slope$\,=\,3.5$,
and (ii) $m_l\,=\,1, m_u\,=\,120,$ slope$\,=\,1.0$. The first one is
heavily biased against high mass stars while the second one is rich in
massive stars. Comparison of the results for the three IMFs is done in
sec.~3.6.

We depict the contribution from different masses to the cluster spectrum as a
function of age of the IB in Figs~3(a)--(f). The mass range
1--60\msun, for the standard IMF is
divided into four sub-intervals, namely (i) low mass stars ($1 < m < 5$), (ii)
intermediate mass stars ($5 < m < 15$), (iii) high mass stars ($15 < m < 40$)
and (iv) very high mass stars ($40 < m < 60$). The quantities chosen are
(a) Bolometric luminosity, (b) Lyman continuum photon rate,
(c)--(f) $B$, $V$, $R$ and $K$ band luminosities respectively. The ordinates
are expressed in units of \ergs\,\msun$^{-1}$ in all figures except (b), where
it is in ph\,s$^{-1}$\,\msun$^{-1}$. The dotted, dashed and dash-dotted lines
represent the contribution from stars in 40--60\msun, 15--40\msun\ and
5--15\msun\ range respectively. The contribution from the 1--5\msun\ bin
always lies below the plotted range. The solid line in the plots is the sum
of contribution from all the stars, which is plotted separately for four of
the optical and NIR bands in Fig.~3(g). The bands are marked
on the curves. Also marked in Fig.~3(g) are the beginning and end of the
supergiant phases for the selected masses. The following conclusions can be
drawn from these figures. \\
1. Early cluster evolution is dominated by the evolution of stars
more massive than 15\msun, with lower mass stars beginning to contribute
substantially at around 10~Myr. \\
2. All quantities except $N_L$ show an initial increase in luminosity
peaking at an age between 5--8~Myr before the decline starts.
The rate of increase is mainly controlled by the evolving massive stars
($>40$~\msun). Peak luminosities are reached earlier at shorter wavelengths
compared to that at longer wavelengths, as the IB cluster evolves. Lyman
continuum photon rate remains steady for the first 3~Myr,
decreasing monotonically then onwards. \\
3. The initial increase in luminosity in optical and NIR bands is
due to the combined effects of the increase in bolometric luminosity
and the decrease in the effective temperature of massive stars as they evolve
off the main-sequence. The increase in luminosity is larger at longer
wavelengths ({\it e.g. } $K$ band).\\
4. Bolometric luminosity and ionizing photon rates are heavily dependent on the
massive luminous stars and hence decrease rapidly once these stars die at
around
5~Myr. On the other hand the luminosities in optical and NIR bands
decrease only marginally after the death of massive stars. \\
5. The peak luminosities in optical and NIR bands during the evolution
of the cluster are due to the red supergiant phase of the 15--40\,\msun\
stars. The red supergiant contribution to total luminosity
as compared to that of the low mass main sequence stars is increasingly
significant at longer wavelengths. For example 99\% of the $K$ band luminosity
is contributed by the red supergiants at around 10~Myr.\\
6. The only regime where masses less than 5\msun\ show their presence is
in the evolution of optical luminosities beyond 12~Myr. Even here their
contribution is less than 20\%. \\
7. Small wiggles in the diagram especially beyond 15~Myr are not genuine, but
are due to the limitations of the computations. \\

The evolutionary results for the three scenarios of star formation are
presented in Tabs~3, 4 and 5. Refer Tab.~2(a) for the explanation of
the column headers in these tables. Tabulated quantities are smoothed over
a period of 0.5~Myr to suppress the short term fluctuations on the
computed quantities. For continuous star formation model, constant SFR is
assumed in the computations of the tabulated quantities. Columns 1 and 2 of
Tab.~5 give the ages of the young and old bursts for the two-burst models.
The data are presented for the younger bursts at ages 0, 3.5, 5.0 and 6.0~Myr
and older bursts of equal strength but evolved to ages between 6 and 14 Myr.
Computational results for only the standard IMF are presented in the printed
form. These Tabs~3--5, containing the results for the three IMFs discussed
above are presented in AAS CD-ROM series.

The Evolution of the most important observable quantities is shown
graphically in Figs~4(a)--(d) for the standard IMF. The solid lines depict the
quantities smoothed over 0.5~Myr with the dotted lines representing the
un-smoothed quantities at intervals of 0.1~Myr for IB scenario. Results for
CSF scenario are plotted as dashed and dot-dash lines corresponding to
exponentially decreasing SFR with $t_0=10$~Myr and constant SFR respectively.
Quantities plotted are $\log$(\philb), $B-V$, $V-R$ and $V-K$ in Figs~4(a),
(b), (c) and (d) respectively. (\philb) drops by two orders of magnitude in
10~Myr in the case of IB scenario. In contrast this quantity falls by only a
factor two during the same time period for CSFR and by a factor of e for ESFR.
For IB scenario, \philb\ follows the evolution of $N_L$, which is controlled
by the massive stars. The continuous creation of massive stars in CSF
models keeps the \philb\ almost steady.

Three distinct phases can be identified in the evolution of optical and
NIR colors for IB. Colors remain constant at their zero age
values for the first 3~Myr, during which period main sequence stars
dominate the cluster evolution. The second phase is characterized by a
red bump during 5--15~Myr. This phase corresponds to the appearance of red
supergiants in the cluster. Colors again become steady at around 16~Myr,
at a value mid-way between the blue colors of main sequence phase and
the average colors of red supergiant phase. During this phase surviving
stars have masses $<$12\,\msun\ and hence the cluster lacks luminous red
supergiants. The color evolution is determined by the slowly evolving
intermediate mass stars and hence colors are steadier. Larger amplitude
of the dotted lines during this phase represents the limitations of the
computations. Increasing the thickness of the mass grid for stars less
massive than 12\,\msun\ is required to reduce these fluctuations.

Evolution of the colors in the red supergiant phase requires a detailed
discussion. $B-V$ attains the reddest value of 0.38 at 7~Myr, drops down
to a value of 0.12 at 9~Myr, shows two small peaks there after
before entering the third phase. $V-R$ shows the first peak at 8~Myr
reaching peak color of 0.4, which drops down to 0.29 at 9~Myr.
Notice that the drop is stronger for $B-V$ compared to $V-R$ and $V-K$.
Infact $V-K$ color becomes reddest at 11.5~Myr. The position of the color
peaks depends on the detailed evolution of massive red supergiants, which,
when present dominate the luminosity in optical and NIR bands. Fig.~3(g) is
useful in understanding the color behaviour.

Colors for CSF remain steady at zero-age values upto 3--4~Myr,
increasing to values of 0.1, 0.2 and 2.0 for $B-V$, $V-R$ and $V-K$
colors respectively. Colors remain steady at these values after about
10~Myr. CSF colors remain bluer than the IB colors during the luminous
red supergiant phase of cluster evolution. Difference between CSFR and
ESFR models is very small compared to their differences from IB models.

\subsection{Cluster Evolution in Observational Plane} 

So far we studied the behaviour of the observable quantities as a function
of age of the cluster. Now we discuss the cluster evolution in $\log$(\philb)
vs color ($B-V, V-R$ and $V-K$) plane. These are shown in Figs~5(a)--(c)
for the standard IMF. Observed quantities can be directly plotted on these
diagrams, and hence the plots are very useful for diagnosis of star formation.
Dotted line denotes the locus of IB model evolved upto 13~Myr, with the thicker
dots placed every 0.5~Myr interval. CSF model with a  constant SFR is shown by
the dashed line, covering 20~Myr of evolution. The solid lines correspond to
models with two bursts of star formation. Four sets of models are plotted.
Age of the younger population is fixed for each set at 0, 3.5, 5 and 6 Myr
from top to bottom respectively. These ages are chosen to represent most
prominent evolutionary phases of the cluster and the corresponding
positions are marked by asterix
on the IB model. The age of the older population varies from 6 to 14~Myr
along a given set, with thick dots denoting 0.5~Myr intervals.

CSF models occupy only a narrow strip in the observational plane even
after evolution upto 20~Myr. On the other hand 2-burst models cover a
much larger range in $\log$(\philb) and colors. It is interesting to
note that these composite models simultaneously produce high \philb\ values
and red colors, when the age of the younger population is $<3$~Myr.
Younger population controls the ordinate, while the colors are
controlled by the red supergiants of the older population. This
phase of cluster evolution can be easily distinguished from IB and CSF
models.

\subsection{Effect of IMF on cluster evolution}  

All the discussions in the previous sections were centered on the
intermediate IMF. In this section we compare those results with two
extreme IMFs in order to judge the effect of IMF differences on observational
quantities. The high mass deficient IMF has parameters $m_l=1$, $m_u=30$
and $\alpha=3.5$, where as the high mass rich IMF has parameters
$m_l=1$, $m_u=120$ and $\alpha=1$.

Evolution of $\log$(\philb), $B-V$, $V-R$ and $V-K$ for IB with the
three IMFs discussed above are shown in Figs~6(a)-(d). Solid line uses
the standard IMF, where as the dotted and dashed lines correspond to
high mass enriched and high mass deficient IMFs respectively. The
initial steady phase in \philb\ and colors prolongs up to 5.5~Myr for the
high mass deficient IMF. \philb\ is lower by an order of magnitude
during this steady phase compared to the standard IMF due to the reduction in
the number of stars contributing to ionizing photons. The red bump in the color
evolution is less pronounced for this high mass truncated IMF. Note that
the colors are redder during the steady phase and bluer in the red
supergiant phase as compared to the standard IMF values.

Evolution with high mass enriched IMF is interesting in several respects.
An examination of Fig.~3, which is for the standard IMF, is useful in
understanding the details of the evolution. The initial steady phase in
\philb\ is almost absent where as it is shortened for colors. This is
due to the domination of optical and NIR luminosities by stars
more massive than 40\,\msun. The extent of domination is more for the
high mass enriched IMF compared to standard IMF. During the early evolution
optical and NIR luminosities increase rapidly where as $N_L$
remains steady resulting in the decrease of \philb.
Due to the shorter lifetime of high mass stars, the cluster gets blue
and red supergiants earlier on during its evolution, leading to shortening
of the initial steady phase in colors. All the stars more massive than
60\,\msun\ cease to exist after 3.5~Myr and hence beyond this age, the high
mass enriched IMF differs from standard IMF only in having a flatter slope.
Evolution of \philb\ is almost independent
of IMFs after 5~Myr. The shape of color evolution during the red
supergiant phase (5--15~Myr) is only weakly dependent on IMF parameters,
but the absolute value of colors during this phase is strongly IMF dependent.
$B-V$, $V-R$ and $V-K$ colors are redder by 0.25, 0.15 and 0.4 mag
respectively in comparison to the standard IMF. Redder colors are due to
an increase of high luminosity red supergiants with respect to low
luminosity red supergiants and main sequence stars, as a result of
the flatter slope. In summary, increasing the fraction of high mass stars
makes the cluster colors bluer in the early phase ($<3$~Myr) and redder during
the red supergiant phase (5--15~Myr).

Effect of IMFs for CSF models are shown in Figs~7(a)--(d). Line types
and notations are as given for Fig.~6. (\philb) has a smooth evolution with
high
mass enriched IMF always having higher value of \philb. Massive star
enriched IMFs have bluer colors during the first 5~Myr and redder colors
beyond that compared to other IMFs. Reddest colors reached by the CSF
models are not as red as that of IB models (0.1 vs 0.65, 0.2 vs
0.55 and 2.0 vs 3.6 for $B-V$, $V-R$ and $V-K$ respectively).

Two-burst model results for the three IMFs are plotted in the
observable planes in Figs~8(a)-(c). The line types used to
distinguish IMFs are as for the previous figures. There are four sets of
models for each IMF. Burst parameters are similar to that for Fig.~5.
Age of the younger population is marked on each track. The high mass
enriched IMF spans a considerable range in the observable plane. The
three IMFs are clearly distinguishable for younger burst ages less than
3.5~Myr. Beyond this age, they are separated only in the optical colors.

\section{Statistical Effects on the Derived Quantities} 

Following the normal approach, we have assumed the stellar masses to be
continuous and have numerically integrated the Eqs~1 and 2 over mass.
Two clusters with identical IMFs can have different distribution of the
stellar masses, leading to different band luminosities, summed over
all the stars. The mean of the luminosities over several clusters with
identical IMFs, is however expected to be the same as that obtained using the
integration approach. The departure from the mean value is dependent on the
cluster mass $M_T$, approaching zero for massive clusters. We investigate here
the minimum mass required for an assumed IMF, such that statistical
fluctuations in star formation does not affect the observable quantities.

We aim to estimate the errors on the computed quantities due to the
statistical selection of stellar masses in a real star forming complex,
as a function of the total mass of the star cluster. This requires a
knowledge of the masses of all the stars in the cluster. Monte Carlo
method is used for simulating real star clusters satisfying the IMF with
$m_l=1$, $m_u=120$ and $\alpha=2.5$. The IMF is normalized such that at
least one star earlier than O4 ($90<m<120$) exists in the cluster. The
mass of the cluster containing only one such star is the minimum mass of
the cluster for the assumed IMF and hence is referred as \Mmin. The
corresponding number of stars in the cluster is \Nmin. For the chosen IMF,
\Mmin$=6641$\msun\ and \Nmin$=2434$. Cluster masses ($M_T$) and total
number of stars ($N_T$) are defined in integer units of \Mmin\ and \Nmin\
respectively. Thus $M_T=n\times$\Mmin\ and $N_T=n\times$\Nmin.
Physically $n$ denotes the number of stars earlier than O4.

For a given mass $M_T$, $N_T$ stars with masses between 1--120\msun\ are
generated weighted by the IMF slope. The slope of the IMF for the
generated mass string is re-determined and confirmed to be within 0.01 of
the input value of 2.5. Luminosity in a given band $B$ from such a mass
string is computed by adding the band luminosities from all the stars.
\begin{equation}
 L_B(imf;t;Z;M_T) = \left( {1\over M_T} \right) \sum\limits_{i=1}^{N_T}
 l_B(m_i;t;Z)
\end{equation}
where $m_i$ is the mass of the $i^{th}$ star generated during the
simulation. $L_B$ is computed at $Z=$\zsun\ for $t=$0--20~Myr at
intervals of 0.1~Myr. The process of generating $N_T$ stars and
computing $L_B$s are repeated for 100 trials, thus simulating 100
clusters with the same IMF and total cluster mass. The error due to the
statistical selection of stars is estimated as rms noise of the 100
trials.

The above exercise is carried out for $n=3,10$ and 50. The mean values
of the computed quantities over 100 trials agree well with the values
obtained from the integration method. Rms errors on the representative
luminosities and colors are plotted in Figs.~9 and 10 for $n=3,10$ and 50.
Errors on the bolometric luminosity, Lyman continuum photon production rate,
expressed in terms of \hb\ flux, $B$ and $K$ band luminosities are plotted
in Fig.~9(a)--(d) respectively. Errors on \philb, $B-V$, $V-R$ and $V-K$
colors are plotted in Fig.~10(a)--(d). Decrease of rms errors with increasing
cluster mass (or increasing $n$) can be easily inferred from these plots.
The errors on colors are
larger during the red supergiant phase of the cluster, as compared to
other phases. For $n=50$ ($M_T=3.3\times 10^5$\msun) the errors on
$\log$(\philb) are less than 0.03 dex throughout the 20~Myr
evolution. Errors remain below 0.03~mag for $B-V$ and $V-R$ colors
during the same period. $V-K$ errors marginally exceed 0.1 mag for ages
greater than 16~Myr. The errors on $\log$(\philb) and colors are larger by a
factor of 2--3 for clusters 5 times less massive ($n=10$). However
these errors are within the various photometric error sources involved
during the photometry of GEHRs (Mayya 1994). Hence for embedded clusters
more massive than $\sim 10^5$\msun\ or more luminous than $10^{39}$~\ergs\
in \ha, the errors due the statistical selection of masses are negligible.
For comparison, 30~Doradus nebula in the Large Magellanic Cloud has \ha\
luminosity of $1.5\times 10^{40}$~\ergs\ (Mathis {\it et al.} 1985).

\section{Comparison with Other Models}   

Currently a number of models exist, which synthesize the cluster parameters
at various wavelength bands. Among these, the most suitable model
for comparison with our results is from Mas-Hesse \& Kunth (1991). We
compare our monochromatic spectrum of a cluster$+$nebula with their table
6(a). Comparisons are made at all the tabulated ages (2, 5, 7.5, 10 and 20~Myr)
for IMFs with $m_l\,=\,2$, $m_u\,=\,120$ and $\alpha\,=\,2.0$. In their model,
Mas-Hesse \& Kunth (1991) have
assumed that only $70\%$ of the available Lyman continuum photons are used up
for ionization. We have taken this into account in our comparison. The results
of the comparisons are given in Figs~11(a)--(e). Spectra integrated over the
entire star forming complex are shown in these Figs, between 70~nm and
3.0~$\mu$m  at solar abundance, for IB model at ages 2, 5, 7.5, 10 and 20~Myr.
Solid line shows cluster spectrum from our computations, whereas the dotted
lines show the total cluster spectrum as computed by Mas-Hesse \& Kunth (1991).
The rms of the differences between the two spectra over the wavelength range
136.2~nm to 2.2~$\mu$m are 0.08, 0.10, 0.18, 0.20 and 0.08~dex for 2, 5, 7.5,
10 and 20~Myr respectively. Hence the agreement is generally good over the
entire wavelength range at all comparison ages except at 7.5 and 10~Myr,
at which time cluster is in its red supergiant phase. Considering that the
spectral shape is rapidly changing during the red supergiant phase and that
Mas-Hesse \& Kunth derived spectrum from available broad band photometry
of red supergiants, the observed differences during this phase are not
un-reasonable.

\section{Summary}    

We have constructed an evolutionary population synthesis model incorporating
the recently available stellar evolutionary models and model stellar
atmospheres. We synthesize a number of observable quantities in the optical and
near infrared part of the spectrum, suitable for comparison with the observed
properties of giant star forming complexes. The ratio of \hb\ to $B$ band
luminosity, $B-V$, $V-R$ and $V-K$ colors have been studied as a function of
cluster evolution for Instantaneous Burst and continuous star formation models.
Computations performed for a two-burst model are also presented with the
younger burst rich in massive stars capable of ionization, and the older burst
rich in red supergiants. Color evolution shows an initial steady phase
for both instantaneous burst and continuous star formation models, followed by
a red bump between 5--15~Myr for instantaneous burst. In the observable
plane, two-burst models occupy quite distinct regions compared to
instantaneous burst and continuous star formation models. Dependence of
computed quantities on the IMF parameters are also studied. Flat IMF
is found to produce the bluest colors when the cluster is young,
and the reddest colors when red supergiants start appearing in the cluster.
The effect of the statistical fluctuations inherent to the process of star
formation on the computed quantities are studied using Monte Carlo
simulations and letting the cluster evolve upto 20~Myr. For GEHRs with
\ha\ luminosities exceeding $10^{39}$~\ergs, the statistical effects are found
to be negligible. The spectra from our models
are compared with the computations of Mas-Hesse \& Kunth (1991) at 2, 5, 7.5,
10 and 20~Myr, and the agreement is found to be reasonably good.

\acknowledgements

It is a pleasure to thank T.P. Prabhu, who took an active interest throughout
the period of this work and contributed greatly through discussions,
suggestions and later by critically reading several versions of the manuscript.
A modelling of this kind is impossible without the availability of input data
in electronic form. I thank G.~Meynet and G.~Schaller for providing
access to their stellar evolutionary data through the network and
M.~Parthasarathy and S.~Giridhar for provided me the data tapes
containing new and old versions of stellar atmospheric models by Kurucz.
I thank the two anonymous referees for their
suggestions which helped in improving the manuscript enormously.

\begin{figure}
\caption{
Lyman continuum photon rate at the stellar surface ($n_{\rm{lyman}}$),
bolometric correction (BC), $B-V, V-R,$ and $V-K$ colors of stars as
computed by us using Schaller {\it et al.}~(1992) and Kurucz (1992) models,
are compared with observations in figures (a), (b), (c), (d) and (e)
respectively as a function of T$_{\rm{eff}}$. Observational data are from
Flower (1977) or Johnson (1966) as indicated, with filled circles
and inverted triangles representing data for main-sequence and
supergiants respectively. The computed quantities for main-sequence and
supergiants are denoted by solid and dashed lines respectively.
Lyman continuum surface fluxes are compared with the computations of
Mas-Hesse \& Kunth (1991). }
\end{figure}

\begin{figure}
\caption{
Equivalent width of \hb\ emission line is plotted against the ratio of \hb\
to blue band luminosity (\philb) in $\log$-$\log$ plot. Age is the
parameter which is varying along the locus, with successive dots separated
by 0.5~Myr. The relationship is not very sensitive to IMF parameters or the
scenario of star formation and hence is useful in converting one form of data
to another.}
\end{figure}

\begin{figure}
\caption{
Contribution to total luminosities from stars in four mass bins are
plotted as a function of cluster age for an instantaneous burst with IMF
parameters $m_l\,=\,1, m_u\,=\,60,$ and slope$\,=\,2.5$. The chosen mass
bins are: $40<m<60$ (dotted line), $15<m<40$ (dashed line), $5<m<15$
(dot-dashed), $m$ expressed in units of \msun. The contribution from the
fourth bin ($1<m<5$) lies below the plotted range. The quantities plotted
are (a) bolometric luminosity, (b) lyman continuum photon rate, (c) $B$
(d) $V$ (e) $R$ and (f) $K$ band luminosities.
The solid lines correspond to the summed contribution from all the stars
which are again plotted in Fig.~(g) for a direct comparison of band
luminosities. The beginning and the end of
the supergiant phases for selected masses are marked on this plot. Ordinates
are in units of \ergs\msun$^{-1}$ for all plots except (b) where it is
ph~s$^{-1}$\msun$^{-1}$. See section 3.4 for further details.}
\end{figure}

\begin{figure}
\caption{
Evolution of (a) \hb\ to $B$ band luminosity ratio ($\log$(\philb)),
(b) $B-V$, (c) $V-R$ and (d) $V-K$ colors
for instantaneous burst and continuous star formation models are shown.
Solid and dotted lines represent smoothed (over 0.5~Myr) and
un-smoothed IB models. Dot-dashed and dashed lines represent CSF models with
constant and exponentially decreasing (e-folding time scale of 10~Myr)
star formation rates respectively. Note the initial steady phase, a red
bump and another steady phase during the IB color evolution.
The IMF used is the same as for Fig.~3.}
\end{figure}

\begin{figure}
\caption{
$\log$(\philb) is plotted against $B-V$, $V-R$ and $V-K$ colors in
figs (a), (b) and (c) respectively. The dotted line is for the instantaneous
burst with thick dots at every 0.5~Myr interval. Solid lines identified by 0,
3.5, 5.0 and 6.0 are tracks for models with co-existing stellar populations
from two bursts of equal strengths. The labeled numbers denote the age of the
younger of the two populations in Myr, whose positions are marked by the
asterix on the IB model. The older population age varies along the track
between 0 and 13~Myr.  CSF model with constant star formation rate is also
plotted (dashed line) for 20~Myr of evolution. The IMF used is the same as for
Fig.~3.}
\end{figure}

\begin{figure}
\caption{
Evolution of (a) $\log$(\philb), (b) $B-V$, (c) $V-R$ and (d) $V-K$ colors
for IB models are shown for 3 IMFs. The parameters of IMF used are:
$m_l=1$, $m_u=120$ and slope$=1$ (dotted line), $m_l=1$, $m_u=60$ and
slope$=2.5$ (solid line) and $m_l=1$, $m_u=30$ and slope$=3.5$ (dashed line).
Notice that as the massive star fraction increases in the cluster, colors
become bluer in the early phase and redder after 5~Myr.}
\end{figure}

\begin{figure}
\caption{
Same as Fig.~6, but for continuous star formation scenario with constant star
formation rate.}
\end{figure}

\begin{figure}
\caption{
$\log$(\philb) is plotted against $B-V$, $V-R$ and $V-K$ colors in
Figs (a), (b) and (c) respectively for the composite two-burst models
for the three IMFs used in Figs~6 and 7. Burst parameters are identical
to Fig.~5. Ages of the younger bursts in Myr are denoted close to the tracks.
Thick dots are placed every 0.5~Myr starting at 6~Myr. IMF
differences stand out for composite models when the younger burst is
younger than 3.5~Myr. Beyond this age differences tend to
decrease especially for $V-K$ color.}
\end{figure}

\begin{figure}
\caption{
Errors due to statistical fluctuations intrinsic to star formation
process on bolometric, \hb, $B$ band and $K$ band luminosities are
plotted for different cluster masses. Numbers on the curves represent
mass of the clusters in units of 6641\msun, the mass containing one star
earlier than O4 for the IMF with $m_l=1$, $m_u=120$ and slope$=2.5$. The
errors are computed as the rms fluctuations on the quantities over 100 Monte
Carlo simulations. See text for details.}
\end{figure}

\begin{figure}
\caption{
Same as Fig.~9, but the errors on $\log$(\philb), $B-V$, $V-R$ and $V-K$
colors are plotted.}
\end{figure}

\begin{figure}
\caption{
Comparison of the cluster$+$nebular spectrum for an IMF with $m_l\,=\,2,
m_u\,=\,120$ and slope$\,=\,2.0$ at ages 2, 5, 7.5, 10 and 20~Myr are
done in Figs~(a),
(b), (c), (d) and (e) respectively for the instantaneous burst model.
Thick lines are from our computations, with dots
corresponding to the wavelengths at which computations are performed. Dotted
lines are from Mas-Hesse \& Kunth (1991). Agreement is good in general
with differences exceeding 0.1~dex only at 7.5 and 10~Myr, at which stage
cluster is dominated by fast evolving red supergiants.}
\end{figure}

\end{document}